\documentclass[preprint2]{aastex}

\usepackage{natbib}

\shorttitle{Bayesian Analysis of Solar Oscillations}
\shortauthors{Marsh, Ireland and Kucera}

\begin{document}

\title{Bayesian Analysis of Solar Oscillations}

\author{M. S. Marsh\altaffilmark{1}, J. Ireland\altaffilmark{2} and T. Kucera}
\affil{NASA Goddard Space Flight Center, Greenbelt, MD 20771, USA.}
\email{mike.s.marsh@gmail.com}

\altaffiltext{1}{NASA Postdoctoral Fellow, NASA/ORAU.}
\altaffiltext{2}{ADNET Systems, Inc.}

\begin{abstract}
A Bayesian probability based approach is applied to the problem of detecting and parameterizing oscillations in the upper solar atmosphere for the first time.
Due to its statistical origin, this method provides a mechanism for determining the number of oscillations present, gives precise estimates of the oscillation parameters with a self-consistent statistical error analysis, and allows the oscillatory model signals to be reconstructed within these errors.

A highly desirable feature of the Bayesian approach is the ability to resolve oscillations with extremely small frequency separations.
The code is applied to SOHO/CDS (Solar and Heliospheric Observatory/Coronal Diagnostic Spectrometer) \ion{O}{5} 629\mbox{\AA} observations and resolves four distinct $P_{4}$, $P_{5}$, $P_{6}$ and $P_{7}$ p-modes within the same sunspot transition region.
This suggests that a spectrum of photospheric p-modes is able to propagate into the upper atmosphere of the Sun and Sun-like stars, and places precise observational constraints on models of umbral eigen modes.
\end{abstract}


\section{Introduction}
The use of Bayesian methods has great potential within astrophysics and has been applied in areas from binary stars to cosmology \cite[see][for an introduction]{lor90}. Bayesian methods have recently been employed within some areas of solar physics, such as the analysis of radiochemical solar neutrino data \citep{stu08}, inversion of Stokes profiles \citep{ase08} and an approach to solar flare prediction \citep{whe04}. The work of \cite{jay87}, \cite{bre88} and others has shown that the application of Bayesian statistical techniques to spectral analysis has many applications in physics, but it has not yet been exploited in solar physics research. Current analysis techniques applied to the problem of wave detection and parameterization in solar physics are not optimal to the problem at hand. Particularly concerning the estimation of oscillation parameters and their uncertainties, it is not clear how to interpret least squares fitting or the Fourier and wavelet transform without understanding a relation to probability theory. It is possible to extract much more information contained within the data by applying Bayesian statistical methods, compared to the traditional least squares, Fourier or wavelet analysis currently employed. The Bayesian method allows extremely precise estimates of the oscillation parameters to be made, with a consistent statistical analysis of their uncertainties. For example, the probability based approach allows the obtainable frequency resolution to be estimated, which is much higher than can be interpreted from a Fourier transform.

We apply the methods described by \cite{jay87} and \cite{bre88} to the problem of frequency estimation within solar data. A Bayesian numerical code is applied to artificial time series data, typical of oscillations within the solar corona, to demonstrate the high precision parameter estimation that can be achieved.
It is shown that frequencies spaced closer than neighboring Fourier frequencies can be successfully resolved by a Bayesian model. This makes the Bayesian approach ideal for determining the number of frequencies present in a time series. Section~\ref{solar} applies the method to transition region data, demonstrating that it is possible to detect and resolve the presence of multiple frequencies in a time series where a Fourier analysis is unable to do so and its application is invalid.

\section{Bayes' Theorem}\label{sect_bayes}
The posterior probability of a hypothesis $H$, given the data $D$ and all other prior information $I$ is stated by Bayes' theorem:
\begin{equation}\label{bayes}
    P(H \vert D,I) = \frac{P(H \vert I)P(D \vert H,I)}{P(D \vert I)}.
\end{equation}
Bayes' theorem derives from commutative logic and the product rule of probability theory \cite[see][]{gre05}.
Where $P(H \vert I)$ is the prior probability of $H$ given $I$, or the prior; $P(D \vert I)$ is the probability of the data given $I$, and is usually taken as a normalizing constant; $P(D \vert H,I)$ is the direct probability of obtaining the data given the hypothesis and prior information. The direct probability is termed the sampling distribution, when the hypothesis is held constant and different sets of data are measured. This sampling distribution has become the traditional approach to estimating the probability of oscillations within astrophysics, particularly within the field of solar physics. However, unlike a laboratory experimenter, or statistician, typically, we can obtain only one measurement of the process under observation. To proceed, the current archetypal method is to assume that the data is one of a large number of possible measurements from a given sampling space. This sampling space is estimated by the application of Monte-Carlo or Fisher-type randomization techniques to generate a large number of artificial `datasets' \citep[e.g. see,][]{osh01, nem85}. Assuming a particular hypothesis, the probability of observing the data within this sampling space of artificial data is then used to estimate the level of confidence in the hypothesis. In the problem of oscillation detection, the level of confidence that there is no oscillating signal present within the data is usually estimated, the null hypothesis.

Since we generally have only one measurement of the data, rather than generating a distribution of artificial `observations', it appears more logical to test the probability of obtaining the measured data against different hypotheses, incorporating the prior information we have available. This is the basis of the Bayesian method. The direct probability is then termed the likelihood function when the data are considered constant and tested against different hypotheses.

\section{Application of Bayes theorem to oscillation detection}
This section summarizes the results of \cite{jay87} and \cite{bre88} which are applied in the code to calculate the Bayesian probability density function and the results in the following sections.
\subsection{The likelihood function}
When applied to the question of oscillation detection, we wish to compute the probability of a particular time series model, given the data and all other prior information. To calculate the likelihood function, the probability of the noise must be calculated. If the true model was known, then the difference between the data and the model function would be equal to the noise distribution. Assuming Gaussian distributed noise, the probability of obtaining a particular series of noise values $e_{i}$ is given by:
\begin{equation}\label{noise}
    P(e_{1}...e_{N} \vert \sigma, I) \propto \prod_{i=1}^{N}\left[\frac{1}{\sqrt{2\pi\sigma^{2}}}exp\left(-\frac{e_{i}^{2}}{2\sigma^{2}}\right)\right],
\end{equation}
where $N$ is the number of elements in the series, and $\sigma^{2}$ is the noise variance. The likelihood function is then given by:
\begin{equation}\label{likely}
    L(\{B\}, \{p\}, \sigma) = \sigma^{-N} exp \{-\frac{1}{2\sigma^{2}} \sum_{i=1}^{N} [d_{i} - f(t_{i})]^{2} \},
\end{equation}
where $d_{i}$ are the measured values of the data. We suppose that the measured data is a combination of the model function and the noise i.e. $$d_{i}=f(t_{i}) + e_{i}.$$ Note that, the data sampling is not required to be evenly spaced, unlike the Fourier transform.

In the most general case, the model as a function of time can be expressed as:
\begin{equation}\label{func}
    f(t)=\sum_{j=1}^{m} B_{j} G_{j}(t, \{p\}),
\end{equation}
where $B_{j}$ are the amplitudes, $m$ is the total number of component model functions $G_{j}$, which are functions of any number of other parameters $\{p\}$, such as frequency, decay rates\ldots etc.
Substituting for $f(t)$, the summation in the likelihood (Eqn~\ref{likely}) becomes:

\begin{equation}\label{Q}
    Q \equiv \bar{d^{2}} - \frac{2}{N} \sum_{j=1}^{m} \sum_{i=1}^{N} B_{j} d_{i} G_{j}(t_{i}) + \frac{1}{N} \sum_{j=1}^{m} \sum_{k=1}^{m} g_{jk} B_{j} B_{k},
\end{equation}
where the cross term for the general model function $f(t)$, in Eqn~\ref{likely}, can be expressed as a matrix of the component model function products summed over time $g_{jk}$.
i.e.

\begin{equation}\label{matrix}
    g_{jk} = \sum_{i=1}^{N} G_{j}(t_{i}) G_{k}(t_{i}), \hspace{0.2in} (1 \le j,k \le m).
\end{equation}

The general model function $f(t)$ in Eqn.~\ref{func} may be composed of any number of component model functions. This is more easily represented in matrix form by a square matrix with indices j,k representing the standard row-major matrix notation. Equation~\ref{Q} is then greatly simplified if the matrix $g_{jk}$ is diagonal.

\subsection{Calculating orthonormal functions}
In the simplest case of an oscillating model function containing a single frequency $f(t)=B_{1}\cos(\omega t) + B_{2}\sin(\omega t)$, the matrix is essentially diagonal due to orthogonality.

In a more complex model containing multiple oscillations, the matrix will not generally be diagonal. To diagonalize the matrix, the component model functions in Eqn.~\ref{matrix} must be transformed to a set of orthogonal functions. The matrix $g_{jk}$ is always a symmetric $m\times m$ square matrix; any matrix of this form has $m$ linearly independent orthonormal eigen vectors and is orthogonally diagonalizable.

The orthonormal model functions are given by:
\begin{equation}\label{h_j}
    H_{j}(t) = \frac{1}{\lambda_{j}} \sum_{k=1}^{m} e_{jk} G_{k}(t),
\end{equation}
where $e_{jk}$ is the $k$th component of the $j$th normalized eigen vector of $g_{jk}$ with a corresponding eigen value $\lambda_{j}$. The functions $H(t)$ then satisfy the orthonormality condition $\sum_{i=1}^{N} H_{j}(t_{i}) H_{k}(t_{i}) = \delta_{jk}$ where $\delta_{jk}$ is the identity matrix. The general model equation (Eqn.~\ref{func}) can then be expressed using these orthonormal component model functions.
The matrix of these functions is then diagonal and Eqn.~\ref{Q} is greatly simplified.

\subsection{Marginalized probability}\label{sect_margin}
The marginalization process allows us to calculate the probability independently of the parameters in which we may have no interest, such as the component model function amplitudes, noise\ldots etc.
Marginalization allows one to remove parameters from further explicit consideration in the posterior distribution, by assigning prior probabilities and integrating the posterior probability distribution over the variable to be removed.
The resulting marginal distribution has no explicit mention of the removed variable, but rather expresses the probability as a function of the remaining variables.
\cite{bre88} derives the probability density as a function of the frequency parameters as follows.

Expressing the summation from the likelihood function (Eqn.~\ref{Q}) using the orthonormal model functions allows the likelihood function to be written in independent terms for each of the component model function amplitudes. The likelihood function is marginalized to be independent of the model amplitudes, by assigning a uniform prior and integrating over each of the amplitudes; this assumes that we have no prior information to constrain the amplitudes of the component model functions. Assuming that we have no prior information to constrain the noise, it can be marginalized in a similar manner to the amplitudes by assigning a Jeffreys prior and integrating over all positive values.
These parameters are marginalized using uninformative priors, where they are not constrained to any particular values. This gives an upper limit to the uncertainty of the parameter estimates. Should we have any prior information to constrain the parameter prior probabilities, then greater precision estimates would be achieved.

This process has a great advantage compared to least squared fitting, in that the probability is evaluated only as a function of the parameters of interest. Thus reducing the dimensionality of the computed parameter space, whereas all parameters must be considered simultaneously using a least squares approach. Even after marginalization of the posterior probability distribution, the Bayesian method still allows good estimates of the marginalized parameters to be recovered without intensive computation, as described in Sect.~\ref{sect_param_est}.

\subsection{The probability density function}\label{sect_pdf}
The resulting posterior probability density that a general oscillatory model is present within the data is given by:
\begin{equation}\label{pdf}
    P\left(\{\omega\}|D,I\right) \propto \left[1 - \frac{m \overline{h^{2}}}{N \overline{d^{2}}}\right]^{\frac{m-N}{2}}.
\end{equation}
This probability density has been derived as a function of the angular frequency parameters only $\{\omega\}$,
assuming data with an unknown noise variance; where $m$ is the number of component model functions, $N$ is the number of measurements in the data time series and $\overline{d^2}$ is the mean square value of the data.

It is the $\overline{h^2}$ function which carries the frequency dependence of the probability density. Given by:

\begin{equation}\label{h2bar}
    \overline{h^2} = \frac{1}{m} \sum_{j=1}^{m}{h_{j}^2},
\end{equation}
where,
\begin{equation}\label{hj}
    h_{j} = \sum_{i=1}^{N}{d_{i} H_{j}(t_{i})}, \hspace{0.2in} (1 \le j \le m).
\end{equation}
The $h_{j}$ values are the projections of the data onto the orthonormal model functions defined by Eqn.~\ref{h_j}, and $\overline{h^2}$ is the mean square value of these projections as a function of $\{\omega\}$.
The maximum of this function gives the most probable frequency $\hat{\omega}$, supported by the data, for each of the component functions assumed by the model. The corresponding maximum in the probability density function (Eqn.~\ref{pdf}) is sharply peaked at these frequency values $\hat{\omega}$, since the form of the function is similar to an exponential. This allows very precise frequency estimates to be made, at a resolution much higher than can be estimated from the Fourier transform, as described in Sect.~\ref{sect_freq_par} and \ref{sect_freq_res}.

In the simplest case where we assume a general model function containing a single stationary harmonic frequency given by $f(t)=B_{1}\cos(\omega t) + B_{2}\sin(\omega t)$, where sine and cosine are the component model functions $G_{j}$ given in Eqn.~\ref{func}, then the eigen values and eigen vectors of the matrix $g_{jk}$ described in Eqn.~\ref{h_j} are equal to $\lambda_{j}=\frac{1}{\sqrt{N/2}}$ and $e_{jk}=\pm \frac{1}{\sqrt{2}}$ respectively. The $\overline{h^2}$ function is the exact general solution; if we approximate by neglecting the negligible non-diagonal elements of the matrix then $\overline{h^2} \equiv{} \frac{1}{N} \mid \sum_{j=1}^{N} d_{j} e^{i \omega t} \mid^{2}$, which is the Schuster periodogram. It is an important, but subtle, point that probability theory shows there is a direct relation between the Schuster periodogram and the probability that there is a single harmonic frequency within the data. As described by \cite{jay87, bre88}, the maximum of the periodogram gives the most probable frequency assuming that: there is a single stationary harmonic frequency present, the value of $N$ is large, there is no constant component or low frequencies, and the data has a white noise distribution. \cite{ire08} take advantage of this fact, by applying an algorithm for automated oscillation detection within the solar corona.

\section{Parameter Estimation}\label{sect_param_est}
Although, in the single frequency case, the periodogram, Fourier transform and $\overline{h^2}$ are similar; $\overline{h^2}$ has been derived from probability theory and can be understood in a statistical sense. This understanding allows estimates of the model parameters, and their precision, to be derived. These estimates cannot be made from least squares fitting, the Fourier transform, or periodogram, alone without understanding their origin in probability theory. Even though the posterior probability density (Eqn.~\ref{pdf}) is derived to be independent of parameters such as the noise variance and model amplitudes, good estimates of these parameters and their uncertainties can be recovered due to the sharpness of the probability density around the most probable frequencies, as described in this section. The parameter uncertainties are not given directly by least squares fitting, or the Fourier transform, which would require a sampling distribution approach. As described in Sect. 2, this is computationally intensive, and it is questionable whether this approach is appropriate given a single measurement of the data. Here we outline estimates of the model parameters and their variance.

\subsection{The expected noise variance $\langle\sigma^2\rangle$}
The Bayesian analysis allows the expectation value of the noise variance within the data to be calculated as:
\begin{equation}\label{sigma2}
	\langle \sigma^2 \rangle = \frac{1}{N-m-2}\left[\sum_{i=1}^{N} d_i^2 - \sum_{j=1}^{m} h_{j}^2 \right].
\end{equation}

The expected noise variance is a function of $\omega$ and is estimated with the $h_{j}$ functions evaluated at the most probable frequencies $\hat{\omega}$ given by the probability density function.
The expectation value of the noise variance is essentially the difference between: the total square value of the data, and the total square value of the data projected onto the orthonormal model functions defined by Eqn.~\ref{h_j}. It is implicit in the Bayesian model that everything within the data that is not fitted by the model is assumed to be noise. Thus the expected noise variance gives an indication to what degree the model represents the data. We may increase the complexity of the model, by the addition of more component model functions. However, once the real signal within the data has been accounted for, the addition of more component functions will have the effect of reducing $\langle \sigma^2 \rangle$ by fitting the noise. A method to determine the point at which the model best represents the true signal is described in Sect.~\ref{best_model}

The percentage accuracy $\epsilon$ of the expected noise variance is given by:
\begin{equation}\label{sigma2_err}
	\epsilon = \sqrt{2/(N-m-4)},
\end{equation}
where $N$ is the number of elements in the series and $m$ is the number of component model functions. The standard deviation accuracy estimate of the expected noise variance is then equal to $\pm \epsilon \sigma^{2}$.

\subsection{The frequency parameters $\{\omega\}$}\label{sect_freq_par}
The most probable frequencies, of the applied model, are evaluated numerically from the location of the maximum within the probability density function described in Sect.~\ref{sect_pdf}. The accuracy of the frequency parameters can be estimated by expanding $\overline{h^2}$ (Eqn.~\ref{h2bar}) in a Taylor series. This accuracy is dependent on the Hessian matrix of $\overline{h^2}$ evaluated at the most probable model frequencies $\hat{\omega}$:

\begin{equation}\label{bjk}
    b_{jk}=-\frac{m}{2} \frac{\partial^{2}\overline{h^2}}{\partial \omega_{j} \partial \omega_{k}}, \hspace{0.2in} (1 \le j,k \le r).
\end{equation}

The estimated angular frequency resolution is given by the variance of the probability density function for $\omega_{k}$:

\begin{equation}\label{delta_omega2}
    \sigma_{\omega_{k}}^{2}= \langle \sigma^2 \rangle \sum_{j=1}^{r} \frac{u_{jk}^{2}}{v_{j}}, \hspace{0.2in} (1 \le k \le r)
\end{equation}
where $u_{jk}$ is the kth component of the jth eigen vector of $b_{jk}$ with a corresponding eigen value $v_{j}$, $r$ is the number of model frequencies, and the expected noise variance $\langle \sigma^2 \rangle$ is evaluated at the most probable frequencies. These most probable frequencies are then equal to $\hat{\omega} \pm \sigma_{\omega_{k}}$.

Equations~\ref{bjk} and \ref{delta_omega2} show that the obtainable frequency resolution is related to how sharply the probability density is peaked around the most probable frequencies and the magnitude of the noise variance within the data. As described in Sect.~\ref{sect_pdf}, the form of the probability density function is sharply peaked around these frequencies; it is this sharpness which allows very high precision frequency estimates to be made, as described in Sect.~\ref{sect_freq_res}, and permits the results obtained in Sect.~\ref{solar}.

\subsection{The amplitude parameters $\langle B \rangle$}\label{sect_amp_par}
If each oscillating function within the model is expressed in the form $f(t)=B_{\cos}\cos(\omega_{} t_{}) + B_{\sin}\sin(\omega_{} t_{}),$ then two component model functions (sine and cosine) are used to describe each frequency component in Eqn.~\ref{func}. Where $B_{sin}$ and $B_{cos}$ represent the amplitude parameters of the sine and cosine functions. For multiple frequency models,
\begin{eqnarray}\nonumber
f(t)=B_{1}\cos(\omega_{1} t_{}) + B_{2}\cos(\omega_{2} t_{}) + \\ \nonumber
B_{3}\sin(\omega_{1} t_{}) + B_{4}\sin(\omega_{2} t_{}) \ldots,
\end{eqnarray}
where the $B_{k}$ parameters of the cosine functions are indexed consecutively for each frequency component, followed by the sine functions.

The expectation value of each amplitude parameter is given by:

\begin{equation}\label{amps}
\langle B_{k} \rangle = \sum_{j=1}^{m} \frac{h_{j}e_{jk}}{\sqrt{\lambda_{j}}}, \hspace{0.2in} (1 \le k \le m),
\end{equation}
where $h_{j}$ are the projections of the data onto the orthonormal model functions in Eqn.~\ref{h_j}, $e_{jk}$ are the components of the normalized eigen vectors of the matrix $g_{jk}$ given in Eqn.~\ref{matrix}, with corresponding eigen values $\lambda_{j}$. These amplitude parameters are dependent on $\omega$, and are estimated using the orthonormal model functions evaluated at the most probable frequency parameters $\hat{\omega}$ described in the previous section. This is a very good approximation, due to the sharpness of the peak in the probability density function which is almost described by a delta-function.

The variance of each amplitude parameter is then given by:
\begin{eqnarray}\label{delta_amp2}
\nonumber    \sigma_{B_{k}}^{2}= \left[ \frac{N}{N-2} \right] \left[\frac{2N-5}{2N-5-2m}\right]  \left[ \frac{2N-7}{2N-7-2m} \right] \\ \left[ \overline{d^2}-\frac{m\overline{h^2}}{N} \right] \sum_{j=1}^{m} \frac{e_{jk}^{2}}{\lambda_{j}},
\end{eqnarray}
where $N$ is the number of time series elements, $\overline{d^{2}}$ is the mean value of the data squared and $\overline{h^{2}}$ is the mean square value of the data projected on to the orthonormal model functions.

\subsection{The polar amplitude $\langle A \rangle$ and phase parameters $\langle \phi \rangle$}\label{sect_phase_par}
The expected amplitudes can be used to express the model function results in polar coordinates, where each oscillation within the model is of the form:
$$f(t)=A\cos(\omega t + \phi).$$ The polar amplitude and phase for each frequency component are then given by: $$\langle A \rangle=\sqrt{B_{\cos}^2+B_{\sin}^2}, \hspace{0.2in}  \langle \phi \rangle=\arctan(-\frac{B_{\sin}}{B_{\cos}}).$$
The $1\sigma$ errors $\sigma_{A}$ and $\sigma_{\phi}$ are then given by the propagation of the $B_{k}$ $1\sigma$ errors given in Eqn.~\ref{delta_amp2}.

\section{Frequency Resolution}\label{sect_freq_res}
In this section, we apply the Bayesian model to artificial test data, typical of the type of oscillations that are observed within the solar corona using current instrumentation.

\subsection{The single frequency case}\label{sect_single_freq}
Here we compare the results obtained from the Bayesian model with those obtained by performing a Fourier, and wavelet analysis. The analyzed time series is of the form:
$$d_{i}= \cos(2\pi f t_{i}) + \sin(2\pi f t_{i}) +e_{i},$$
where $f$=3.3~mHz and $e_{i}$ is Gaussian distributed noise. This is typical of the 5-minute period band oscillations observed in coronal loops, with 100 samples of 30~s cadence and has a low signal to noise ratio with a RMS S/N=1.  The Bayesian model is applied to calculate the probability density function for a single harmonic frequency model of the time series, using Eqn.~\ref{pdf}. The most probable frequency, given by the peak within the probability density function, can be described by a Gaussian of width $\sigma_{\omega}$ given by Eqn.~\ref{delta_omega2}, converting from angular frequency to give $\sigma_{f}$ in Hz. This width determines the theoretical frequency resolution obtainable with the Bayesian model, which is much less than the estimated frequency resolution obtained using the Fourier transform.

\begin{figure}[t]
\centering
\plotone{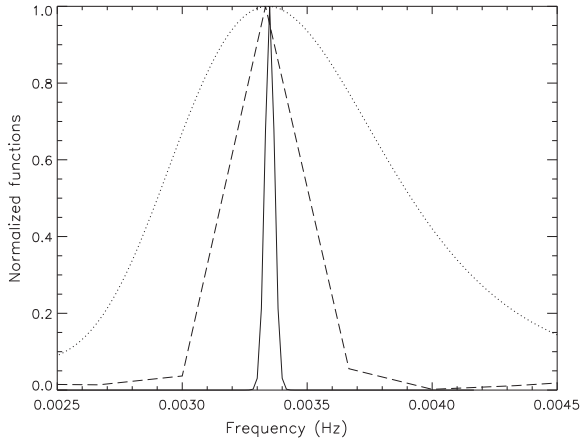}
\caption{A comparison of the frequency resolution obtained for the single frequency data with a S/N=1, with the normalized Bayesian probability density function (solid line), FFT (dashed) and global wavelet transform (dotted). We see that the Bayesian PDF has an order of magnitude increase in resolution compared to the FFT.}
\label{pdf_fft_wav}
\end{figure}

\begin{table}[t]
\centering
\caption{Obtained frequency resolutions for the single frequency data with a  S/N=1.}\label{tab1}
\begin{tabular}{ll} \hline
Analysis method & $\delta f$ (mHz) \\ \hline
Bayesian PDF ($\sigma_{f}$) & 0.02 \\
FFT (HWHM) & 0.2 \\
Wavelet (HWHM) & 0.5 \\
\hline
\end{tabular}
\end{table}

Figure~\ref{pdf_fft_wav} shows the obtainable resolving power of the Bayesian model, normalized to that obtained by the Fourier and wavelet transforms. The solid line indicates the probability density function for the single frequency Bayesian model, the dashed line shows the FFT and the dotted line is the global wavelet transform using the Morlet wavelet function. We can see that the Bayesian model gives a significant increase in the resolution of the estimated frequency, with the probability density almost described by a delta-function. Table~\ref{tab1} lists the obtained frequency resolution for each method, estimating the resolution of the Fourier and global wavelet transforms using their half width half maximum (HWHM). We see that, for a S/N=1, the $1\sigma$ error on the estimated frequency from the Bayesian model gives an order of magnitude increase in resolution over the FFT. The global wavelet has an even lower resolution due to the smoothing effect of the transform on the wavelet scale. In fact, if we are interested in high precision frequency measurements of stationary frequencies, or closely separated frequencies, then a wavelet analysis is one of the worst methods that we can apply.

\subsection{Two closely separated frequencies}\label{sect_2_close_freq}
\begin{figure}
\centering
\plotone{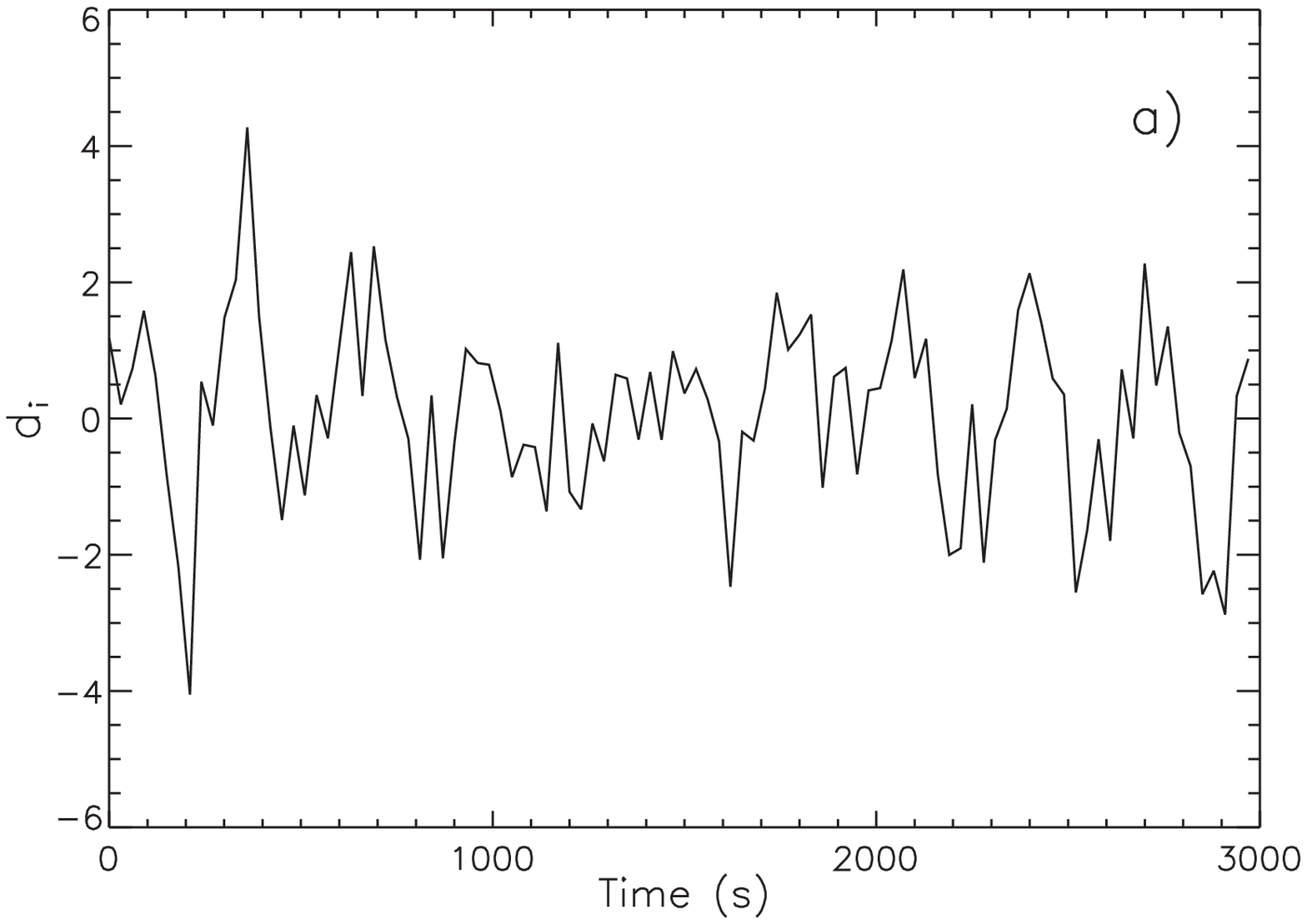}
\plotone{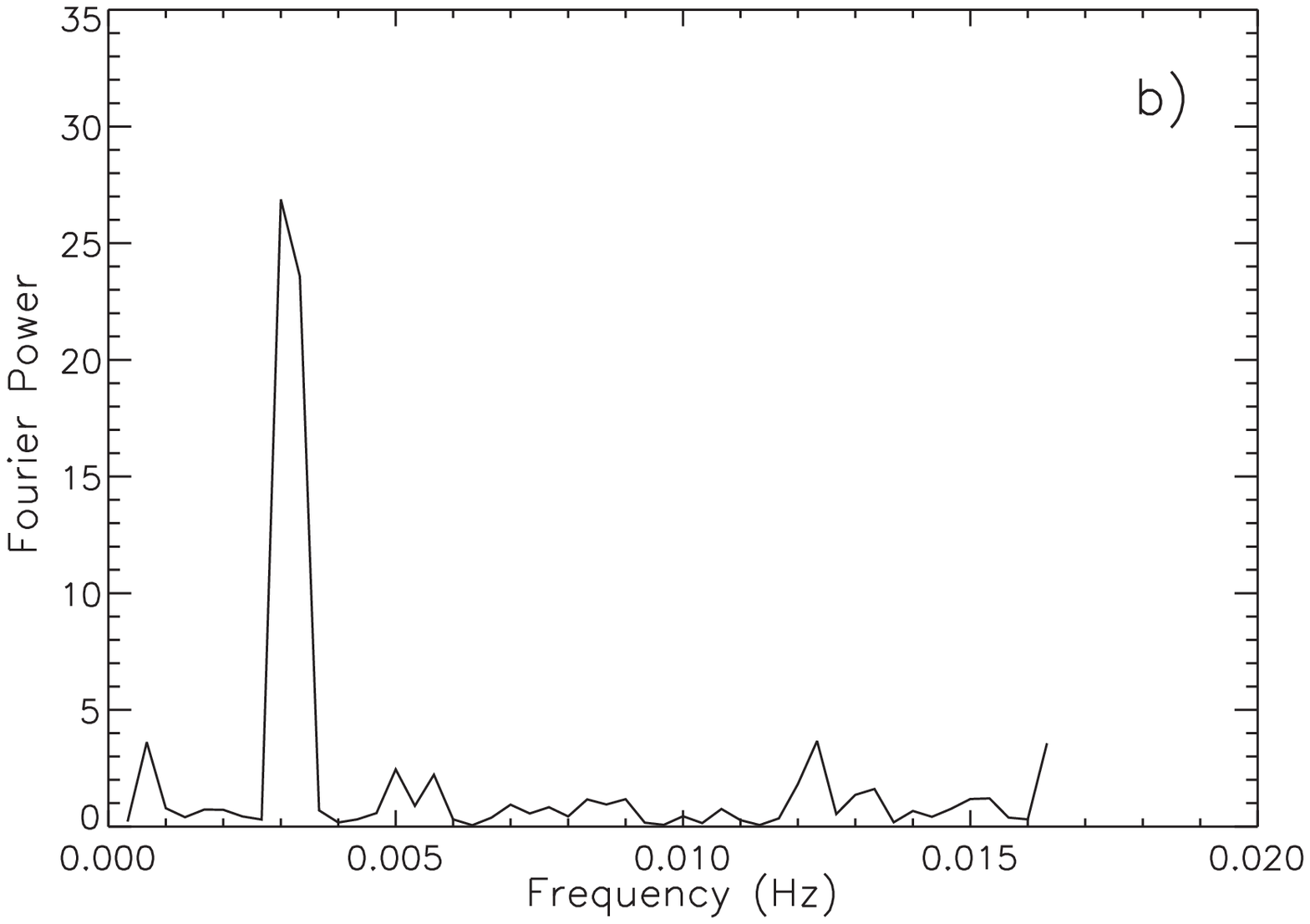}
\plotone{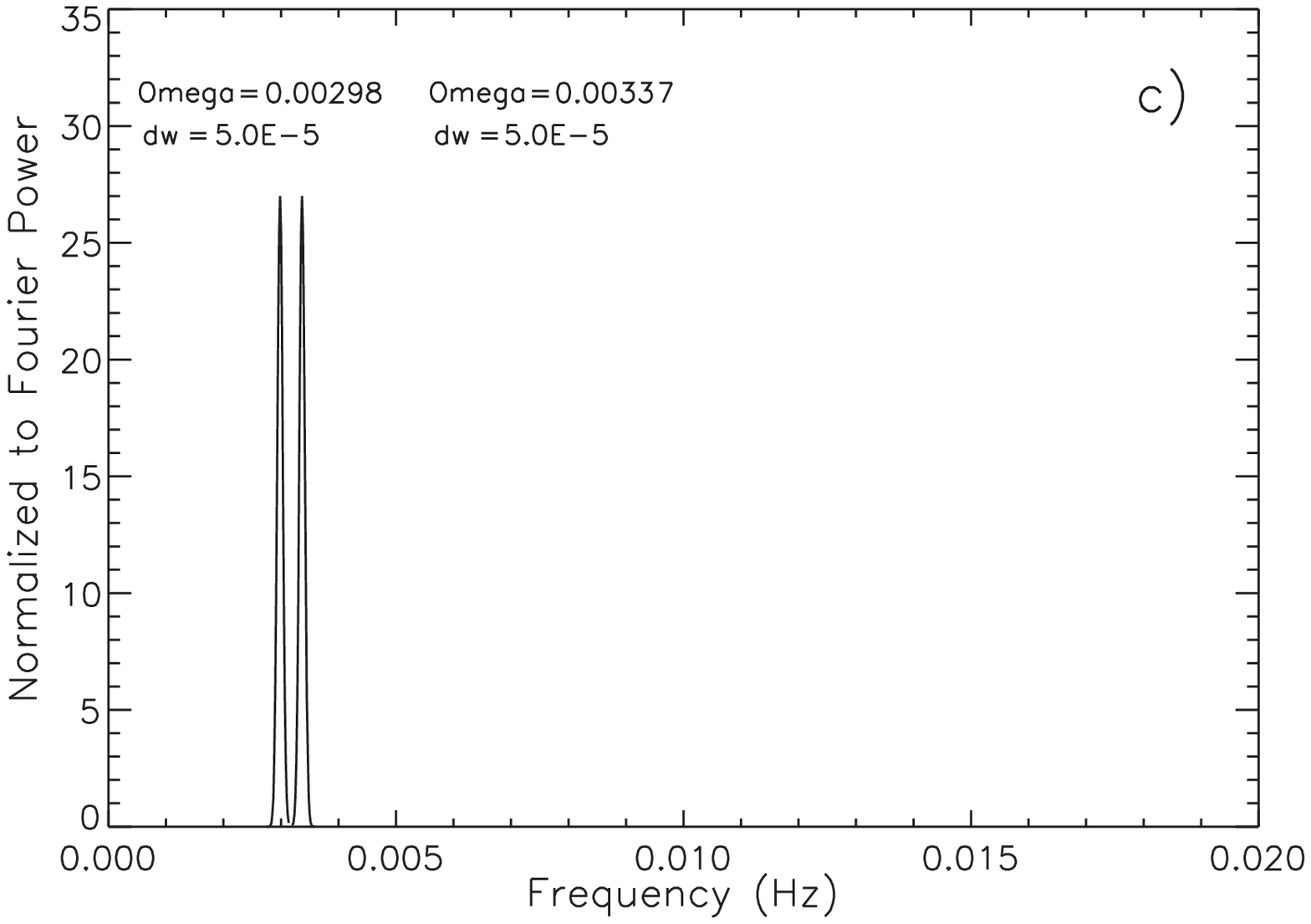}
\plotone{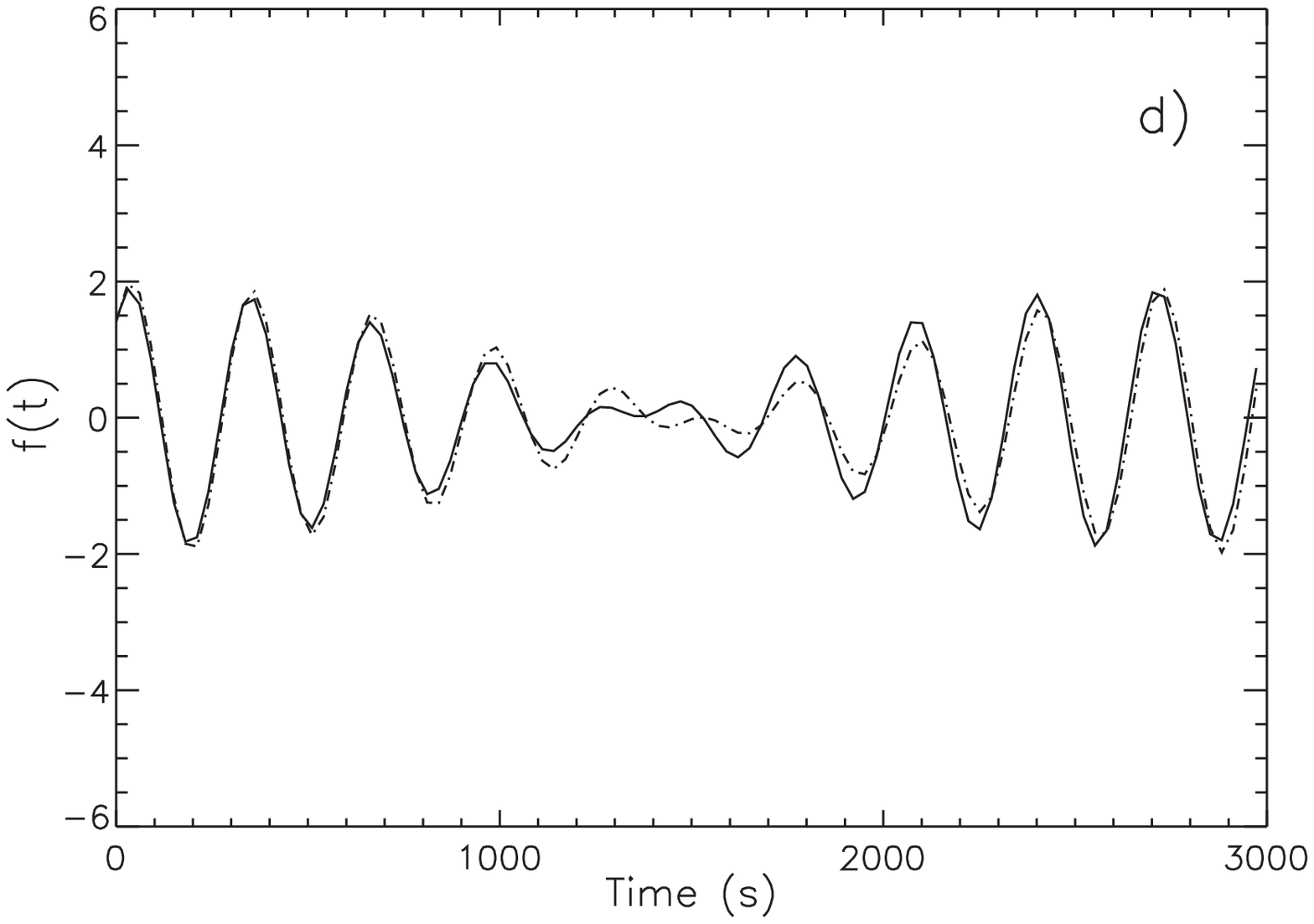}
\caption{a) Two frequency time series with a S/N=1. b) FFT of the time series containing two frequencies separated by 1 Fourier frequency step. c) Gaussian representations of the frequency resolution obtained from the Bayesian probability density function, normalized to the FFT peak. d) Signal reconstructed from the Bayesian model (solid), true signal (dot-dash).}
\label{2freq}
\end{figure}

We now compare the results obtained from a Bayesian model and a Fourier analysis of two closely separated frequencies within a simulated time series shown in Fig~\ref{2freq}a. Again, we generate a time series typical of coronal loop oscillations of the form:
\begin{eqnarray}
\nonumber d_{i}=B_{1}\cos(2\pi f_{1} t_{i}) + B_{2}\cos(2\pi f_{2} t_{i}) + \\ \nonumber
B_{3}\sin(2\pi f_{1} t_{i}) + B_{4}\sin(2\pi f_{2} t_{i}) + e_{i},
\end{eqnarray}
with 100 samples at 30~s cadence, a harmonic frequency of $f_{1}$=3.3~mHz, an additional frequency of $f_{2}$=3.0~mHz, Gaussian distributed noise $e_{i}$ and a RMS S/N=1. These two frequencies are separated by only one frequency step in the Fourier transform, so in principle their frequencies are directly adjacent in the FFT.

\begin{table}[t]
\centering
\caption{Estimated frequencies and $1\sigma$ errors, of the parameters expressed in polar coordinates, obtained from the Bayesian model applied to the two frequency time series, separated by one Fourier frequency step and a RMS S/N=1.}\label{tab2}
\begin{tabular}{cc}
\tableline\tableline
Frequencies $f \pm \sigma_{f}$ (mHz) & True value (mHz) \\ \tableline
2.98 $\pm$ 0.05 & 3.00 \\
3.37 $\pm$ 0.05 & 3.33 \\
\tableline
Amplitude $A \pm \sigma_{A}$ & True value \\ \tableline
0.96 $\pm$ 0.19 & 1.00 \\
0.94 $\pm$ 0.19 & 1.00 \\ \tableline
Phase $\phi \pm \sigma_{\phi}$ (rad) & True value \\ \tableline
5.64 $\pm$ 0.14 & 5.50 \\
5.52 $\pm$ 0.15 & 5.50 \\ \tableline
$\langle\sigma^2\rangle$ & True value \\ \tableline
0.88 $\pm$ 0.13 & 0.89 \\ \tableline
\end{tabular}
\end{table}

Figure~\ref{2freq}b shows the FFT for the two frequency time series; Fig.~\ref{2freq}c shows the Gaussian representation of the resolution obtained from the Bayesian probability density function, which has been normalized to the FFT peak for comparison. Note Fig~\ref{2freq}c is not a power spectrum, but an illustration of the frequency resolution obtained with the Bayesian model. As expected the FFT is unable to resolve two such closely separated frequencies. A single broad peak is observed, with a large HWHM, suggesting the possibility that more than one frequency may be present. However, the result from the Bayesian model resolves the two frequencies independently and to a very high precision even with a S/N=1. Table~\ref{tab2} lists the resolved frequencies and their $1\sigma$ errors, estimated from the probability density function of the two harmonic frequency Bayesian model. We see that the Bayesian model not only resolves the two frequencies but does so to a very high precision with 1$\sigma_{f}$ errors of 0.05~mHz, even with a relatively short duration time series. Figure~\ref{2freq}d shows the signal reconstructed from the Bayesian model parameters, and the true signal within the simulated time series. We see that the Bayesian code provides a very good reconstruction of the signal even with a low signal to noise ratio.

\section{Application to Solar Data}\label{solar}
The Bayesian model is now applied to real observations of solar oscillations. \cite{me06} observe the apparent propagation of slow-magnetoacoustic waves within a sunspot region. These waves are observed to propagate from the transition region into the coronal loop system emerging from the sunspot and are interpreted as the propagation of photospheric p-modes waveguided along the magnetic field.

The original analysis applied Fourier techniques to the time series; here we apply the Bayesian model to the \ion{O}{5} data described in \cite{me06}. The results presented in \cite{me06} show the presence of two frequencies, in the 3-min period band, observed in the transition region above the sunspot umbra. The data consist of 100 samples observed at 26~s cadence obtained with the Coronal Diagnostic Spectrometer (CDS) \citep[see, ][]{har95}. The Bayesian model is applied iteratively, with the addition of further model functions to increase the complexity of the applied model.

\subsection{Model selection}\label{best_model}
In addition to the problem of fitting a model to the data, we must determine which is the most probable model of those under test. If we already have a knowledge of the noise variance within the data, then the calculated expectation value of the noise variance (Eqn~\ref{sigma2}) may be used to determine when the model has accounted for the real signal. Component functions may be added to the model until the expectation value of the noise variance equals the known noise variance. At this point the most appropriate model has been determined, and the addition of further component functions to the model will simply be fitting the noise.

\begin{figure}[t]
\centering
\plotone{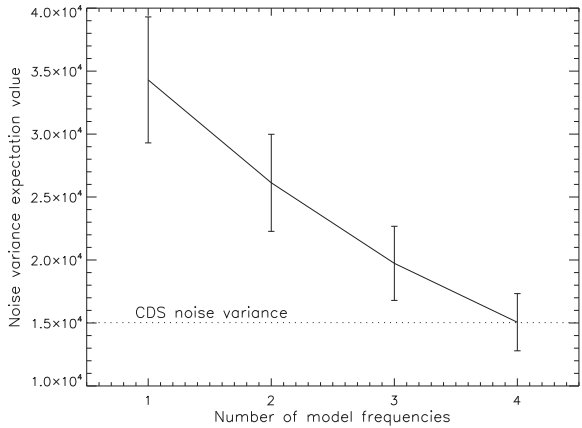}
\caption{Change in the expectation value of the noise variance for an increasing number of component frequencies within the Bayesian model. The dotted line indicates the known noise variance within the CDS data.}
\label{model_var}
\end{figure}

The noise properties of the CDS instrument are described by \cite{sn49}. The noise variance of the CDS detector is given by $\sigma^{2} = 2N_{ph} + R^{2}n$, where $N_{ph}$ is the number of detected photons, $R$ is the readout noise (here we use a conservative value of 1 photon-event pixel$^{-1}$), and $n$ is the number of pixels summed over. We may use this known value of the noise variance to determine when the model has reached sufficient complexity to account for the real signal and the addition of further model functions will begin to fit the noise.

Figure~\ref{model_var} shows the change in the expectation value of the noise variance for increasingly complex models with the addition of more component frequencies. The error bars show the standard deviation error estimate of the expected noise variance derived from Eqn.~\ref{sigma2_err}. The dotted line shows the noise variance within the data due to the photon statistics and the noise properties of the CDS detector. As expected, with the addition of further component frequencies to the model, the expected noise variance is reduced. The expected noise variance reaches the level of the CDS data with a model containing four harmonic frequencies. Therefore we can state that the data best supports a model containing four frequencies. We are able to derive parameter estimates of these functions and their associated errors, from the probability density function of the four frequency model.

\subsection{The Bayesian results}
\begin{figure}
\centering
\plotone{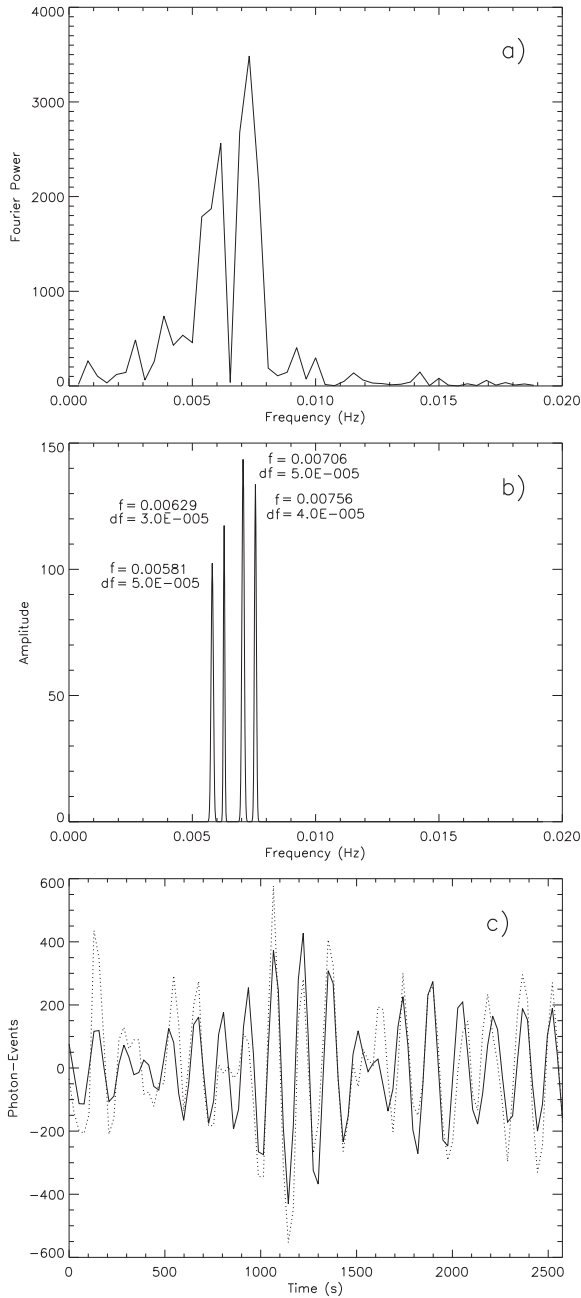}
\caption{a) The FFT of the \ion{O}{5} data originally presented in \cite{me06} showing two frequencies in the 3-min range. b) Gaussian representations of the frequency resolution obtained from the Bayesian probability density function for the four frequency model, where the peaks are normalized to the derived amplitudes given in Table~\ref{tab3}. c) Signal reconstructed from the Bayesian model (solid), original data (dot-dash). }
\label{ov_fft_4freq}
\end{figure}

Figure~\ref{ov_fft_4freq}a shows the FFT of the \ion{O}{5} data presented in \cite{me06}, with two main peaks resolved in the transform. The broad width of the peaks may suggest that the data does not simply consist of two monochromatic frequencies, but this is the limit to the information available using a Fourier analysis. Fig.~\ref{ov_fft_4freq}b shows the Gaussian representation of the four frequencies resolved by the Bayesian model. In addition to the oscillation frequency, the Bayesian code also returns high precision estimates of the amplitude and phase parameters, listed in Table~\ref{tab3}. The height of the peaks in Fig.~\ref{ov_fft_4freq} are normalized to the corresponding amplitude parameters; note that this figure is not a power spectrum, but a representation of the frequency resolution obtained from the probability density function. As demonstrated in Sect.~\ref{sect_2_close_freq}, the Bayesian model is able to resolve closely spaced frequencies to a much higher resolution than is possible with the Fourier transform, even with short duration observations.  Where the FFT can resolve only two frequencies, the Bayesian model is able to resolve four independent frequencies within the data, to a very high precision.

\section{Conclusions}
When approaching the problem of oscillation detection, it is possible to extract much more information from the data by the application of a Bayesian model rather than the traditional least squares fitting, Fourier, or wavelet analysis. Considering the problem of frequency estimation within a time series, the Bayesian method returns very precise estimates and employs a rigorous self-consistent error analysis, due to its statistical derivation. It is not clear how to determine frequency error estimates from the Fourier transform or least squares, without understanding their relation to probability theory. It is often misconceived that the Fourier frequency spacing is the limit to the resolution with which a frequency can be resolved within a time series. This is not the case, as the resolution limit is principally determined by the signal to noise ratio. As shown in Sect.~\ref{sect_single_freq}, we are able to estimate a single frequency, with S/N=1, to a resolution an order of magnitude greater than the FFT. In Sect.~\ref{sect_2_close_freq}, we are able to resolve two oscillations with frequency separations directly adjacent in the FFT.  This is not surprising as the Bayesian method has similarities with least squares fitting of a particular function. In an analogous way, the limiting resolution with which an oscillation can be fitted with a sinusoidal function is not equal to the time series cadence, nor is the resolution with which a spectral line can be fitted with a Gaussian equal to the pixel spacing on the detector; this resolution limit is largely determined by the signal to noise ratio within the data.

\begin{deluxetable}{ccc}
\tabletypesize{\scriptsize}
\tablecaption{Bayesian model parameters derived from the \cite{me06} data.\label{tab3}}
\tablewidth{0pt}
\tablehead{
\colhead{Frequencies $f \pm \sigma_{f}$ (mHz)} & \colhead{Amplitude $A \pm \sigma_{A}$ (Photon-Events)} & \colhead{Phase $\phi \pm \sigma_{\phi}$ (rad)}
}
\startdata
5.81 $\pm$ 0.05 & 102.4 $\pm$ 26.2 & 0.10 $\pm$ 0.18 \\
6.29 $\pm$ 0.03 & 117.3 $\pm$ 26.1 & 1.35 $\pm$ 0.16 \\
7.06 $\pm$ 0.05 & 143.6 $\pm$ 26.2 & 2.63 $\pm$ 0.13 \\
7.56 $\pm$ 0.04 & 133.6 $\pm$ 26.3 & 5.31 $\pm$ 0.14 \\ \tableline
Expected noise variance $\langle\sigma^2\rangle$ & 15062.0 $\pm$ 2271.0  & \\
\tablenotemark{a}CDS noise variance $\sigma^2$ & 15029.0 & \\
\enddata
\tablenotetext{a}{CDS noise variance calculated using \cite{sn49}}

\end{deluxetable}

\cite{lou78} describe how the Fourier transform gives erroneous results for closely spaced frequencies. Their results are also explained by probability theory; \cite{jay87} discovered that the periodogram or Fourier transform is only directly related to the probability of a single stationary harmonic frequency within the data. If multiple frequencies are well separated, the Fourier transform still gives good frequency estimates, as the problem separates out into independent single frequency probability problems. If the frequencies are closely spaced, however, the non-diagonal elements in the matrix $g_{jk}$ (Eqn.~\ref{matrix}) become significant and the frequencies are not orthogonal. The approximation of using a single frequency probability model for the purpose of frequency estimation is no longer valid, nor is the use of the Fourier transform. In this case, the transformation to orthogonal functions used in the Bayesian analysis is necessary to determine accurate frequency estimates and their uncertainties.

As mentioned by \cite{bre88}, the Bayesian method is similar to least squares, in that a least squares approach minimizes the summation in Eqn.~\ref{likely}, whereas the Bayesian results maximize the likelihood function. As described in Sect~\ref{sect_margin}, the Bayesian method allows only the parameters of interest to be considered, greatly reducing the dimensionality of the parameter space compared to a least squares approach. In the previous section, a least squares approach would require a 12 dimensional parameter space, where this is reduced to 4 dimensions with the Bayesian model. In principal, least squares will produce similar frequency estimates to the Bayesian model, since uninformative priors have been used, however, least squares does not directly determine their uncertainties. The Bayesian framework allows the oscillatory model to be understood in terms of probability theory, and achieves higher precision estimates due to the sharp maximum of the posterior probability density.

We apply the Bayesian model to the \ion{O}{5} data presented in \cite{me06}, and resolve four closely spaced independent frequencies within the 3-minute period range. The observations presented in \cite{me06} are interpreted as the conversion and propagation of photospheric p-mode oscillations along the magnetic field into the corona. 
5-minute period range oscillations are observed within the umbral photosphere of sunspots, and are shown to be connected to the global p-mode oscillation distribution centered on 5-minutes \citep{pen93, bal87, bra87}. Oscillations in the 3-minute period range have been observed in the chromosphere above sunspot umbrae for many years \citep{bec69, bec72, gur82, lit82}. The 3-minute period range oscillations are thought to be due to amplitude steepening of the photospheric p-mode spectrum \citep{bog00}. Recent work by \cite{cen06} supports this, demonstrating that 3-minute range power in the chromosphere is due to linear wave propagation from the 5-minute range power in the photosphere. The 3-minute oscillations are also observed in the umbral transition region \citep{tho87,flu01, osh02, ren03, bry04}.
The results presented here suggest that we are able to resolve these oscillations into four closely spaced p-mode frequencies. \cite{zhu02} calculate the spectrum of eigen modes within the vertical magnetic field of the sunspot umbra, finding that the 3-minute umbral oscillations are due to p-modes modified by the strong magnetic field within the sunspot. \cite{zhu05} calculates the same spectrum of umbral oscillations using the method of resonant filtering and by solving the eigen value problem, also determining that the 3-minute oscillations are part of the photospheric p-mode spectrum which propagates through the umbral atmosphere. The frequencies detected here, and their spacing, are consistent with the model results of \cite{zhu05} and may represent the detection of the $P_{4}$, $P_{5}$, $P_{6}$ and $P_{7}$ photospheric p-modes in the solar transition region. These results provide precise observational constraints for future modeling of umbral eigen modes. A more detailed discussion, on the characterization of these resolved modes, is required than can be addressed here. A following paper will investigate these modes and their constraints on a model atmosphere.

%
%

\acknowledgments
M.S. Marsh is supported by the NASA/ORAU postdoctoral program, and would like to acknowledge the encouragement of L.E. Pickard. SOHO is a project of international cooperation between ESA and NASA. We would like to thank the anonymous referee for useful comments that improved the manuscript.

{\it Facilities:} \facility{SOHO (CDS)}.

\bibliographystyle{apj}
\bibliography{ms2col}

\end{document}